\documentclass[preprint]{aastex}

\begin{document}

\title{Spectral Irradiance Calibration in the Infrared. XIV: the Absolute Calibration of 2MASS}

\author{Martin Cohen}
\affil{Radio Astronomy Laboratory, 601 Campbell Hall, University of
California, Berkeley, California 94720\\
Electronic Mail: mcohen@astro.berkeley.edu}

\author{Wm. A. Wheaton}
\affil{California Institute of Technology, Infrared Processing \& Analysis
Center, MS 100-22, 770 South Wilson Avenue, Pasadena, CA 91125\\
Electronic Mail: waw@ipac.caltech.edu}

\author{S. T. Megeath}
\affil{Harvard-Smithsonian Center for Astrophysics, 60 Garden Street MS-65,
Cambridge, MA 02138\\
Electronic Mail: tmegeath@cfa.harvard.edu}

\begin{abstract}
Element-by-element we have combined the optical components in the three
2MASS cameras, and incorporated detector quantum efficiency curves and
site-specific atmospheric transmissions, to create three relative spectral
response curves (RSRs).  We provide the absolute 2MASS attributes associated
with ``zero magnitude" in the $JHK_{s}$ bands so that these RSRs may be used
for synthetic photometry. The RSRs tie 2MASS to the
``Cohen-Walker-Witteborn" framework of absolute photometry and stellar
spectra for the purpose of using 2MASS data to support the development of
absolute calibrators for IRAC and pairwise cross-calibrators between all
three SIRTF instruments. We examine the robustness of these RSRs to
changes in water vapor within a night. We compare the observed 2MASS 
magnitudes of thirty three stars (converted from the 
precision optical calibrators of Landolt and Carter-Meadows 
into absolute infrared (IR) calibrators from 1.2--35~$\mu$m) 
with our predictions, thereby deriving
2MASS ``zero point offsets" from the ensemble. These offsets are the final
ingredients essential to merge 2MASS $JHK_{s}$ data with our other
absolutely calibrated bands and stellar spectra, and to support the
creation of faint calibration stars for SIRTF.
\end{abstract}

\section{Introduction}

In an ongoing series of papers, Cohen and his colleagues have described a
framework for absolute IR calibration that embraces a variety of spaceborne,
airborne, and ground-based photometers and spectrometers (see Cohen et al.
(1999: hereafter ``Paper X"), and references therein). The framework
currently consists of an all-sky network of over 600 stars (Walker \& Cohen 
2002), each represented
by a complete, absolute, low-resolution spectrum from 1.2 to 35~$\mu$m. This reliance
on calibrated spectra provides the flexibility to incorporate any
well-characterized photometer's passbands, and spectrometers, into this
common calibration scheme. Cohen et al. (2003: hereafter ``Paper XIII") have
described the extension of this approach from purely IR template spectra to
``supertemplates" and Kurucz (1993) models, that extend from the ultraviolet
to the mid-IR. These procedures were designed to furnish on-orbit absolute
calibrators for the instruments on board NASA's Space InfraRed Telescope
Facility (SIRTF), notably for the Infrared Array Camera (IRAC).

The technique involves normalizing a spectral supertemplate or model shape,
appropriate to the spectral type and luminosity class of a star and suitably
reddened, on the basis of optical photometry of the individual star. IRAC has
four detector arrays, with central wavelengths of 3.6, 4.5, 6.5 and 8.0~$\mu$m. 
The IRAC detectors are roughly 2000 times more sensitive than those of
IRAS and would have saturated on all but the faintest stars of the existing
network. To create faint calibrators for IRAC, 2MASS $JHK_{s}$ data will be
used to support these normalizations in the near-infrared (NIR). Therefore,
it is essential to incorporate the 2MASS RSRs into the larger context
described in Papers X and XIII, which already include approximately 110 
characterized passbands.

The present paper describes our characterization of the three 2MASS
passbands, incorporating information on the optics of the 2MASS cameras, their
detector properties, the atmosphere above the two 2MASS telescopes and, in
particular, the robustness of the 2MASS RSRs to variations in precipitable
water vapor above the telescopes. \S2 describes the elements that contribute
to the RSRs of the three cameras, and tabulates the requisite combinations of
optics+filter+detector+atmosphere that represent the path of starlight
through each camera. \S3 presents the RSRs, and provides their
absolute ``zero magnitude" attributes consistently with all other
characterized passbands already within Table 2 of Paper X (IR bands), and 
Tables 3 and 12 of Paper XIII (optical RSRs). \S4 investigates
the influence of variations in water vapor on the calibration of the $J$-band. 
\S5 builds on the set of thirty three absolute calibrators of intermediate
brightness recently constructed in Paper XIII, 
by predicting their 2MASS magnitudes and comparing these with those
actually observed. These comparisons are used to determine the ``zero point
offsets" (hereafter ZPOs: see \S5 for their definition) for 2MASS, 
the final prerequisites before one is
able to normalize supertemplate and model spectra directly by 2MASS
photometry (see Table 3 of Paper X for details of the ZPOs for other
IR photometric systems).

\section{The 2MASS instruments}
\subsection{Optics}
\begin{figure}[tbp]
\centering
\vspace{18cm} \includegraphics{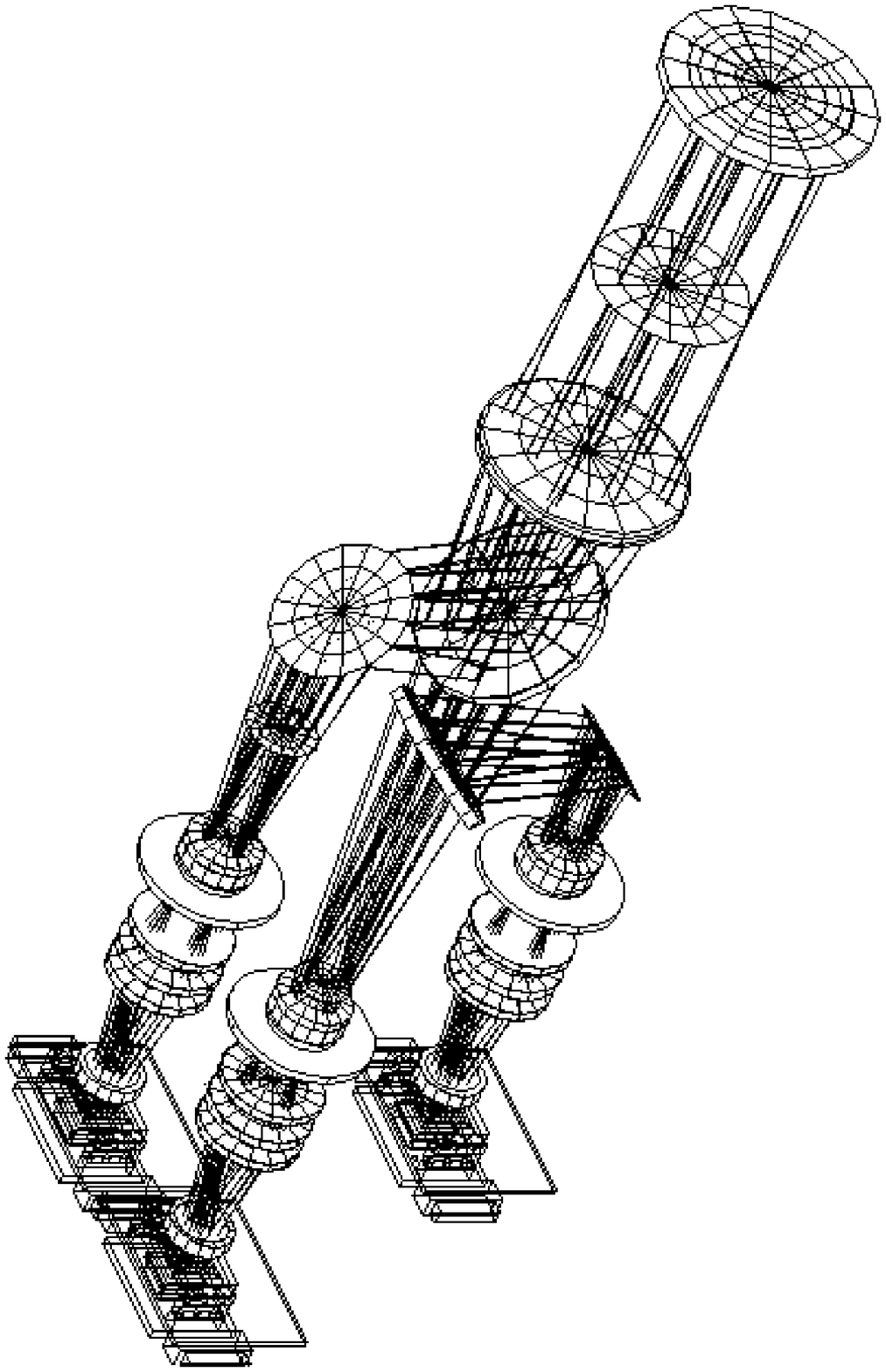}
\caption{The optical trains of the 2MASS cameras.  After the two reflections in 
the telescope, the first camera element, proceeding from right to left, is the 
dewar window, followed by a field stop and the first of the seven lenses. 
Corresponding lenses are identical among the three cameras. Following the first 
lens (the only one common to all cameras) are the two dichroic mirrors, first $J$ 
and then $H$. The straight-through light path leads to the $K_{s}$ camera, the 
upper path to the $J$ camera, and the lower one to $H$.}
\label{camera}
\end{figure}

Our effort to characterize the end-to-end RSRs of 2MASS is based on the
transmission data for the various components of the cameras 
\footnote{http://www.ipac.caltech.edu/2mass/releases/allsky/doc/sec3\_1b1.html},
which appears as part of the Explanatory Supplement for
2MASS\footnote{http://www.ipac.caltech.edu/2mass/releases/allsky/doc/explsup.html}.
Figure~\ref{camera}\footnote{http://www.ipac.caltech.edu/2mass/releases/allsky/doc/figures/seciii1bf2.gif}
illustrates light paths through the overall instrument, while
Table~\ref{instrument} arranges these components from star to detector for
each 2MASS band. No end-to-end wavelength-dependent RSRs were measured for the
2MASS system.  Therefore, we have constructed these, from the products of the reflection and
transmission characteristics of the relevant elements in the optical trains, to yield
single files that can be used for synthetic photometry.
 
\begin{table}[tbp]
\dummytable\label{instrument}
\end{table}

The primary and secondary mirrors of the pair of 2MASS telescopes have identical 
coatings.  The plotted curve at the explanatory supplement's 
URL is for a single reflection
and, therefore, appears squared when one accounts for total system transmission.
The transmission of the external dewar window applies identically to the $JHK_{s}$ paths.
All three optical paths through the cameras contain seven coated lenses:
one common lens ahead of the dichroics and six lenses per channel behind the 
dichroics.  The ``Camera Lens Coatings" curve on the Web is for transmission through a 
single lens element with its coatings.  Thus, it is raised to the 7th power 
to assess the total spectral transmission through each lens assembly.  
The three dichroic mirrors occur in different combinations of reflection and transmission
to provide the distinct light paths for the cameras (Table~\ref{instrument}).  
The $J$, $H$, and $K_s$ filter profiles quoted were tested at 77~K.
The optical coatings were almost certainly tested at room temperature but the
vendor provides no specifications or comparisons to document their behavior at
the actual operating temperatures in the 2MASS cameras.  The appropriate
curves are already flat so that the expected wavelength shifts and shape changes
are probably insignificant.  The same consideration applies to the dichroic 
profiles: steep gradients in transmission and/or reflection occur in regions 
already blocked by the filters.

\subsection{The NICMOS3 arrays}
\begin{figure}[tbp]
\centering
\vspace{20cm} \includegraphics{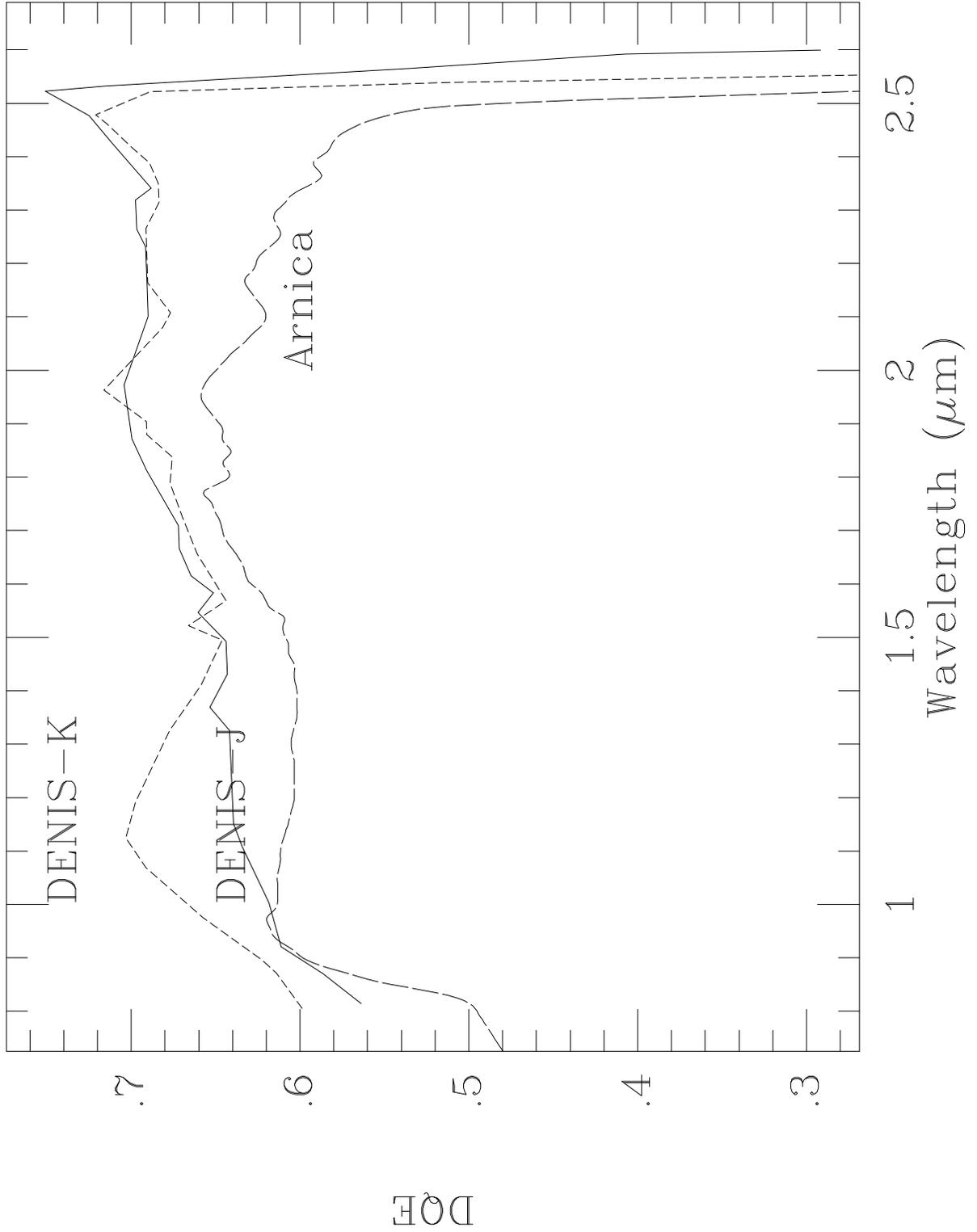}
\caption{The DQE curves for the actual DENIS $J$ and $K_{s}$ NICMOS3 arrays,
and the generic curve supplied with ARNICA.}
\label{dqe}
\end{figure}

Adequate data exist within the 2MASS project
for all components except for the detector quantum efficiency (hereafter
DQE) curves that pertain to the actual 2MASS
detector arrays.  Only limited work was performed by Rockwell, in general, to
characterize DQE as a function of wavelength.  No specific characterization
was provided for the 2MASS arrays.  However, real curves were measured and supplied
by Rockwell as part of the characterization and calibration of 
DENIS (Fouqu\'{e}~ et al. 2000) for 12 regions
surrounding the DENIS detectors, in each of the boules that 
provided the material from which the DENIS $JK_{s}$ detectors were cut.  
The Arcetri Near-Infrared Camera 
(ARNICA\footnote{http://www.arcetri.astro.it/irlab/instr/arnica/arnica.html}) 
also contains a 256$\times$256 NICMOS3
array and was tested in Tucson at the same time as the 2MASS arrays were
tested.  Dr. L. K. Hunt has kindly provided us with the generic DQE information
that she was offered by Rockwell when ARNICA's array was delivered, and she
reports that ARNICA has lower quantum efficiency than the 2MASS devices.
Figure~\ref{dqe} compares the actual DQE curves for the DENIS $J$ and $K_{s}$
arrays with the generic information accompanying ARNICA.  There are broad
similarities between the structure displayed in the curves for all three 
devices, particularly within the range from 1--2.2~$\mu$m.
We have, therefore, chosen to represent the 2MASS DQE curves by the ``generic"
curve accompanying ARNICA, which was intended as an exemplar of the material
used in the 256$\times$256 format.
The generic curve clearly satisfies the characteristics of NICMOS3 HgCdTe,
as suggested by the DENIS curves.  Given our ignorance of the actual 2MASS 
devices' performance with wavelength, the qualitative statement 
that the overall value of the DQE for 2MASS's arrays exceeds that for ARNICA 
is not an issue because we will renormalize all derived RSRs so their peaks are
unity.  We are grateful to Dr. M. F. Skrutskie for locating a set of plots 
from Rockwell characterizing the HgCdTe material produced 
at about the same time that the 2MASS NICMOS3 arrays were in production (1996).  
This documents DQE at 8 locations, albeit for a 1024$\times$1024 array.  The
curves are in broad agreement with those we offer in Figure~\ref{dqe}, and
combine the broad plateau seen in ARNICA around 1.3~$\mu$m with the 
elevated levels of the DENIS arrays beyond 2.0~$\mu$m.  

\subsection{The contribution of the atmosphere}
To represent the intervening telluric transmissions above Mt Hopkins, AZ (MHO) and
Cerro Tololo Inter-American Observatory (CTIO) in Chile, we represented the atmospheric
transmissions using PLEXUS\footnote{http://www2.bc.edu/\~\,sullivab/soft/plexus.html\#Desc}, 
an AFRL validated ``expert system" that incorporates atmospheric code, specifically 
MODTRAN 3.7, SAMM, and FASCODE3P with the HITRAN98 archives.
PLEXUS contains an extensive database to support its expert aspect, so that
the effects of aerosols and particulates appropriate to the desert conditions 
of the 2MASS sites were included.  Paper X also used PLEXUS calculations 
to represent the site-specific atmospheric transmissions necessary to represent
the more than one hundred ground-based filters characterized in that work.

We first verified that the ratio of PLEXUS-computed transmissions at MHO and CTIO 
was flat over the entire passbands of the $H$ and $K_{s}$ filters, and almost
all the $J$ band, and very similar.  Therefore, the use of a single calculation to 
incorporate the atmosphere for both 2MASS telescopes is justified.

\subsection{Assembling the RSRs}
With the sole exception of the DQE curves, every element in the 2MASS instrument 
is represented by data germane to the final survey cameras.  Strictly, the direct 
products of all these component curves are valid only if we can ignore the
``Stierwalt Effect".  If an interference filter is placed in close proximity to 
a detector array, the out-of-band-rejection may be greatly changed due to the 
increased solid angle seen by the detector. Filters are usually measured in a
``collimated" beam, the rejected flux being scattered into a larger solid angle. 
We have explicitly assumed that this effect is not an issue within the 2MASS cameras.
Lacking quantitative uncertainties for the characteristics of all these components, 
we have assigned an overall wavelength-independent error of 5\% to the resulting RSRs.

\section{2MASS RSRs}
\begin{figure}[tbp]
\centering
\vspace{20cm} \includegraphics{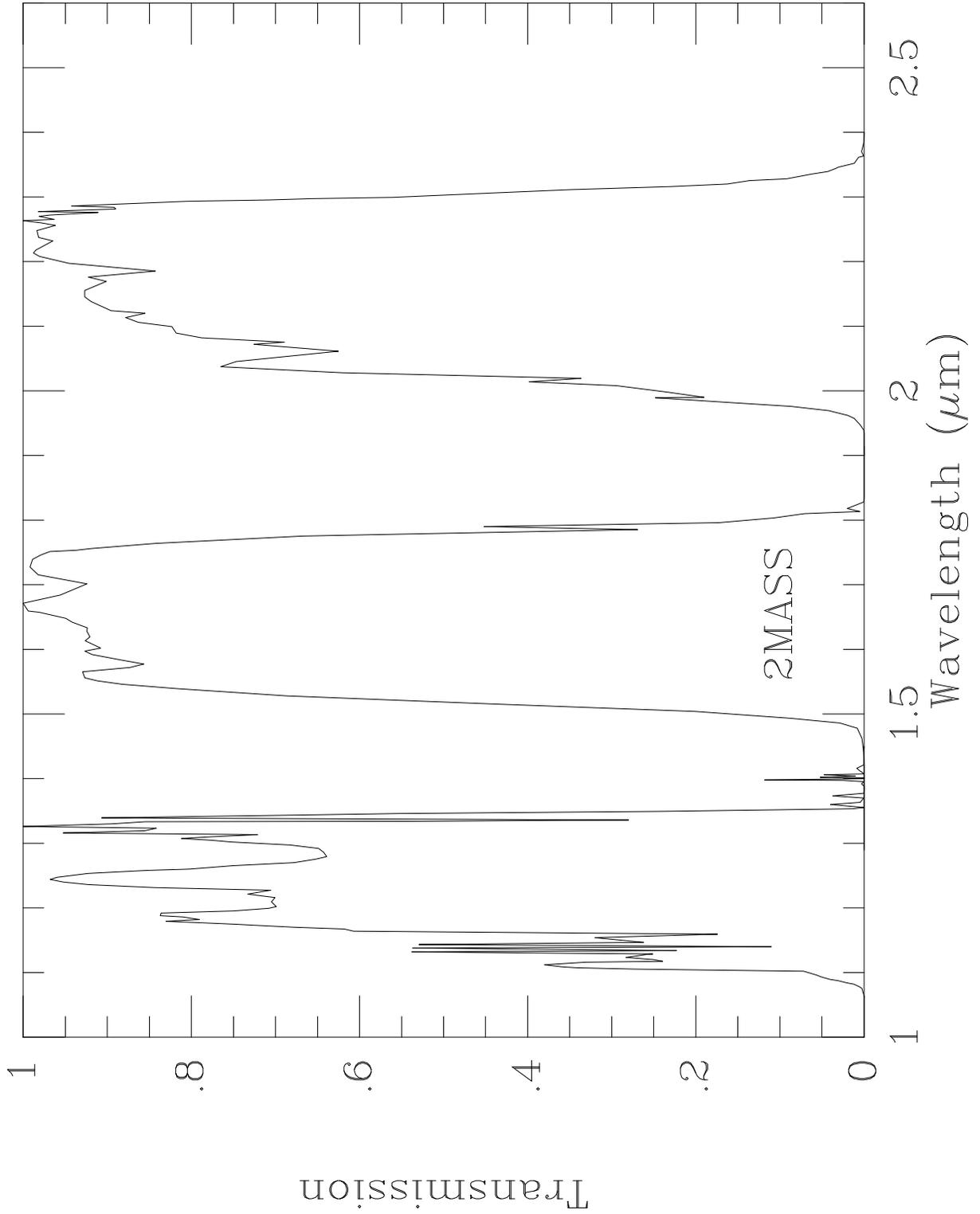}
\caption{The calculated relative spectral response curves for 2MASS bands,
renormalized to peak values of unity.}
\label{rsrs}
\end{figure}

Figure~\ref{rsrs} presents the three RSRs for 2MASS, including all the optics
described above, the (ARNICA) generic DQE curve, and a single PLEXUS atmosphere representative
of typical survey conditions.  We have followed standard practice by
normalizing each RSR to unity at its peak.  Note that these RSRs are designed
to be integrated directly over stellar spectra in F$_\lambda$ form, in 
order to calculate synthetic photometric magnitudes.  The quantum efficiency
based component was converted to yield photon-counting RSRs by multiplying 
by $\lambda$ and renormalizing, exactly as described by Bessell (2000).
These three RSRs can be found on the Web at:
%http://www.ipac.caltech.edu/2mass/releases/allsky/doc/figures/seciii1b1f21.gif ($J$);
http://www.ipac.caltech.edu/2mass/releases/second/doc/sec3\_1b1.tbl12.html ($J$);\\
http://www.ipac.caltech.edu/2mass/releases/second/doc/sec3\_1b1.tbl13.html ($H$);\\
http://www.ipac.caltech.edu/2mass/releases/second/doc/sec3\_1b1.tbl14.html ($K_s$),
but will soon be moved to analogous URLs on the ``allsky" site.

\subsection{The influence of water vapor variations}
\begin{figure}[tbp]
\centering
\vspace{20cm} \includegraphics{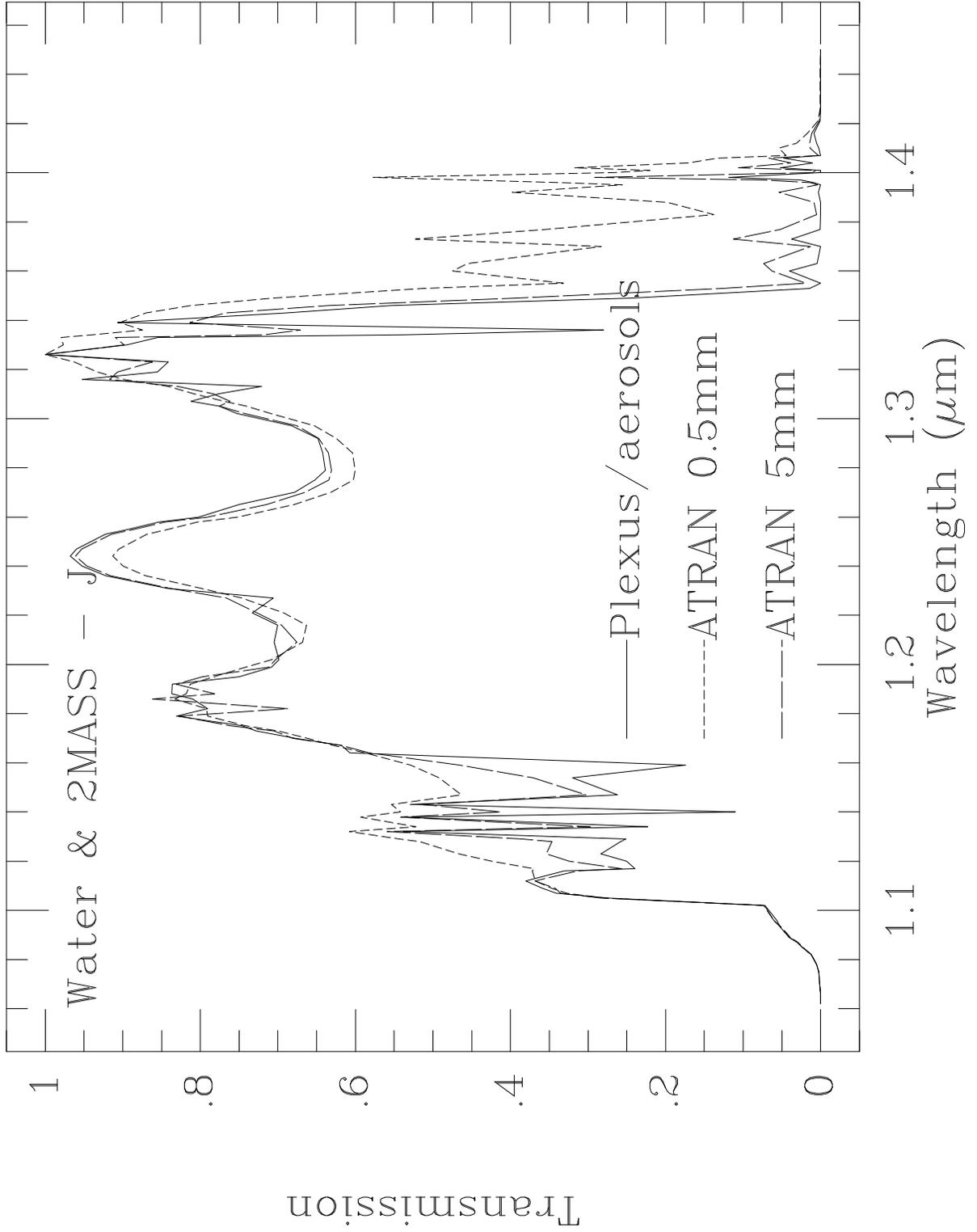}
\caption{Changes in the 2MASS $J$ RSR due to variations in water vapor.}
\label{jh2o}
\end{figure}

On some 2MASS survey nights, as much as a 20\% variation in response 
in the $J$-band channel has been seen, presumably due to changes in telluric 
transmission.  Figure~\ref{rsrs} indicates that the $J$-band RSR is significantly 
influenced by water vapor near 1.35~$\mu$m.  It is particularly susceptible to 
intra-night, as well as night-to-night, variations. 
If all else were unchanged, it is of interest to find out how variable the 
response might be if it depended solely on the water vapor content of the atmosphere,
especially in the $J$-band. Therefore, we investigated the RSRs that would
arise as a result of an order of magnitude variation in precipitable water,
specifically for amounts of 0.5 and 5~mm. The former would have been a
rather rare occurrence during 2MASS operations, while somewhat more than 5~mm would 
represent the worst conditions under which survey data were collected. 
These calculations were performed with the ATRAN code (Lord 1992).  Calculating for
the altitude of MHO, ATRAN predicts about 6.5~mm of precipitable water, but ATRAN
enables arbitrary changes, to our selected values of 0.5 and subsequently 
to 5~mm, without altering
any other aspect of the atmosphere (note that, although ATRAN lacks aerosols,
particulates, and an expert database, it is a highly flexible code for the study of
water vapor variations).  Figure~\ref{jh2o} illustrates the pair of resulting RSRs.
The final adopted $J$ RSR also
appears in Figure~\ref{jh2o} where it is seen to compare very favorably with
the ATRAN transmission curve for 5~mm of water. We still prefer the PLEXUS products
because they have been used throughout Paper X, and for DENIS (Fouqu\'{e}~et
al. 2000), and because the ``expert" mode offers additional realism
to our characterization of the atmosphere.  The amount of precipitable water 
vapor in an ATRAN model is not directly comparable to that implied by a PLEXUS model
for the same site because of the dependence on the database of actual measurements 
in PLEXUS, and the inclusion of other phenomena.

No effects were found on the 2MASS $H$ or $K_{s}$ bands due to these gross 
variations in water vapor.

\section{The absolute calibration of 2MASS}

To intermingle optical and IR photometry in the normalization of calibrated
supertemplates (Paper XIII) requires that every band be
well-characterized and that all are integrated over the identical, absolutely
calibrated, Kurucz spectrum of Vega published by Cohen et al. (1992), which
underpins the entire context of Paper X and, consequently, that of 2MASS and
SIRTF too. In the IR we adopt the synthetic Vega spectrum as the definition
of zero magnitude, whereas small but nonzero magnitudes apply in the optical
(e.g. Bessell et al. 1998; Paper XIII).

Table~\ref{cal} details the attributes for zero magnitude for 2MASS, giving
in-band fluxes, isophotal wavelengths and frequencies, and the corresponding
isophotal monochromatic intensities in F$_{\lambda}$ and F$_{\nu}$ units, all
calculated using the RSRs in this paper.
2MASS $JHK_{s}$ can thereby be transparently compared with photometry in any
of the bands in Paper X, and with spectra from all instruments using
this common calibration scheme that unifies ground-based, airborne, and
spaceborne calibrators. Table~\ref{water} offers an estimate of the {\it differences}
in the absolute calibration of the 2MASS $J$-band for both 0.5 and 5~mm of water, 
based on the results of constructing RSRs using ATRAN, not PLEXUS.  Although
the resulting in-band fluxes vary by almost 11\%, the isophotal quantities
are quite robust in showing only a 2\% variation. 

Our {\it recommended} calibration of 2MASS at $J$ appears in Table~\ref{cal}, 
based on PLEXUS.  Table~\ref{water} is
shown purely to quantify the impact of isolated water variations on the $J$-band
absolute calibration.

\begin{table}[tbp]
\dummytable\label{cal}
\end{table}

\begin{table}[tbp]
\dummytable\label{water}
\end{table}

\section{Zero point offsets}

2MASS represents an extremely large body of system magnitudes, defined with
respect to an internally homogeneous set of reference stars. Because these
stars are not already members of the all-sky network of relatively bright
calibrators of Paper X, we require one final step before we can convert
2MASS magnitudes into physical units that are self-consistent with those used by
the Diffuse InfraRed Background Explorer, InfraRed Telescope in Space, Kuiper
Airborne Observatory, Midcourse Space eXperiment, Infrared Space Observatory, 
and SIRTF. This step is the definition of the 2MASS
ZPOs. The ZPOs are necessary in order to compute absolute quantities using
2MASS magnitudes and Table~\ref{cal}, in the form 
$10^{-0.4({\it m}+{\it z})}\times~F_{\lambda}$(0$^{m}$), where {\it m} is the 
observed 2MASS magnitude and {\it z} is the algebraic ZPO for that band. 
Ideally, any photometric system calibrated in the common context of Paper X would 
be based on the use of any
of the more than 600 stars now represented by complete, absolute, 1.2--35~$\mu$m 
spectra (e.g., those of Paper X). However, the extension of this network to a
set of calibrators of intermediate brightness (i.e. $K$=4--10), suitable for
use by 2MASS, has occurred only very recently, in readiness for SIRTF.
Therefore, we will compute the ZPOs {\it post facto}, as the ensemble-averaged 
algebraic differences between observed and predicted 2MASS $JHK_{s}$
for a set of stars previously absolutely calibrated using other
well-characterized NIR data not based on 2MASS.

Paper XIII sets out the procedures by which the suite of IRAC absolute
calibrators has been defined, and then applies them to the creation of a set of
thirty three 1.2--35~$\mu$m fiducial K0-M0IIIs and A0-A5V stars, drawn from two
sets of precision standards: Landolt's (1992) optical standards, and the Carter 
\& Meadows (1995) optical-NIR photometric stars.  These stars now extend
the all-sky network described in Paper X downward by factors of 100-1000 in
IR brightness. At these levels, they are suitable for cross-ties to 2MASS.
The twenty four cool giants are based on $BVRI$ data (supplemented when available by
Hipparcos and Tycho data, also calibrated in the common context: see Table 12 in
Paper XIII) and on
well-characterized $JHK$ measurements either from Tenerife (``TCS") or South Africa 
(``SAAO": see Paper X, Tables 2 and 3). The nine A-dwarfs are based entirely on $BVRIJHK$ 
data from Carter \& Meadows (1995).

By integrating these absolute spectra through the newly-defined 2MASS
RSRs and converting the resulting in-band fluxes into magnitudes using Table~\ref{cal}, 
we have derived the predicted set of 2MASS magnitudes.  
Direct comparison of these with
the final version of observed 2MASS magnitudes enables us to define the 
``mean algebraic deviation" (hereafter MAD) of a star, by averaging the $JHK_s$ 
differences for that star.  Table~\ref{cml} presents the data for all thirty three
stars, giving, for every star: name; spectral type; predicted and observed 2MASS $JHK_{s}$ 
magnitudes and uncertainties; the differences between these,
with their associated errors; and the MAD.  The ensemble average MAD results from
combining all the (observed-minus-predicted) differences among the thirty three stars,
either without weighting (-0.001$\pm$0.005 mag) or using inverse-variance 
weighting (+0.002$\pm$0.003 mag).  We conclude that this set of stars has no 
significant bias, rendering it ideal to define the 2MASS ZPOs.

The uncertainty used for each observed 2MASS magnitude is the associated $j$, $h$, or 
``$k\_msigcom$", i.e. the ``complete" error which incorporates the results of processing 
the photometry, internal errors (from $\surd$N photon statistics and sky background), 
and calibration errors (nightly zero point, flat fielding, and normalization
uncertainties for each band).
We were able to use differences for 33 $J$, 30 $H$, and 32 $K_{s}$
magnitudes. $H$-band data for three stars (HD172651 = SA110-471; SA107-35;
SA108-827) were rejected because of saturation and/or the concomitant
reduced ``N out of M" (where only N measurements with aperture photometry
above 3$\sigma$ were obtained out of a possible M, a common occurrence for
objects on the R1 saturation threshold), or the intrusion of an artifact
on one star's $H$-band image. One $K_{s}$ magnitude (for SA110-471)
was rejected from consideration because fewer than 3 non-saturated frames
were obtained (``ph\_qual"=``E"). All rejected magnitudes and the associated 
MADs are identified by an asterisk in Table~\ref{cml}.

\begin{figure}[tbp]
\centering
\vspace{20cm} \includegraphics{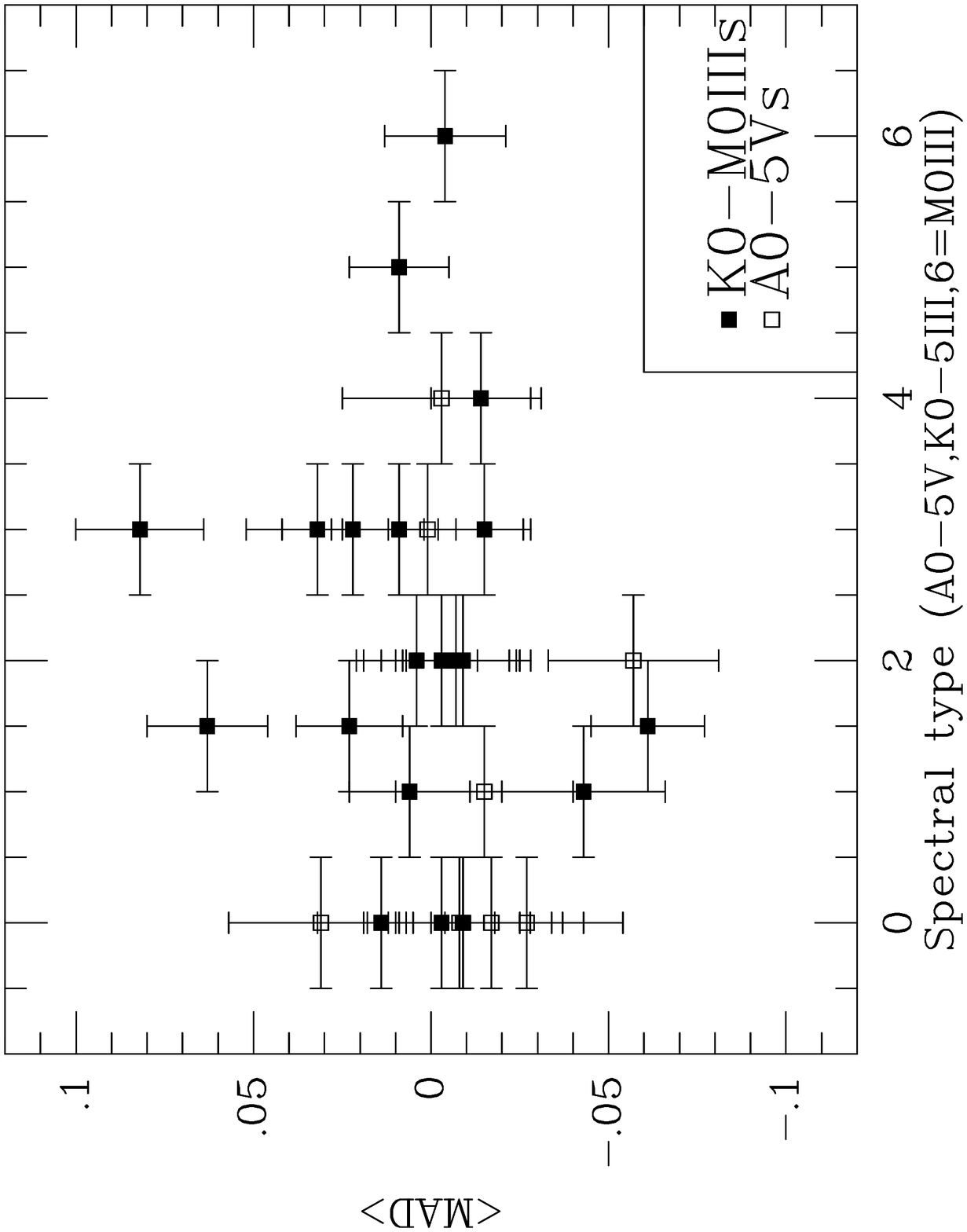}
\caption{MADs for the ensemble of 33 stars as a function of spectral type.}
\label{madspt}
\end{figure}

\begin{figure}[tbp]
\centering
\vspace{20cm} \includegraphics{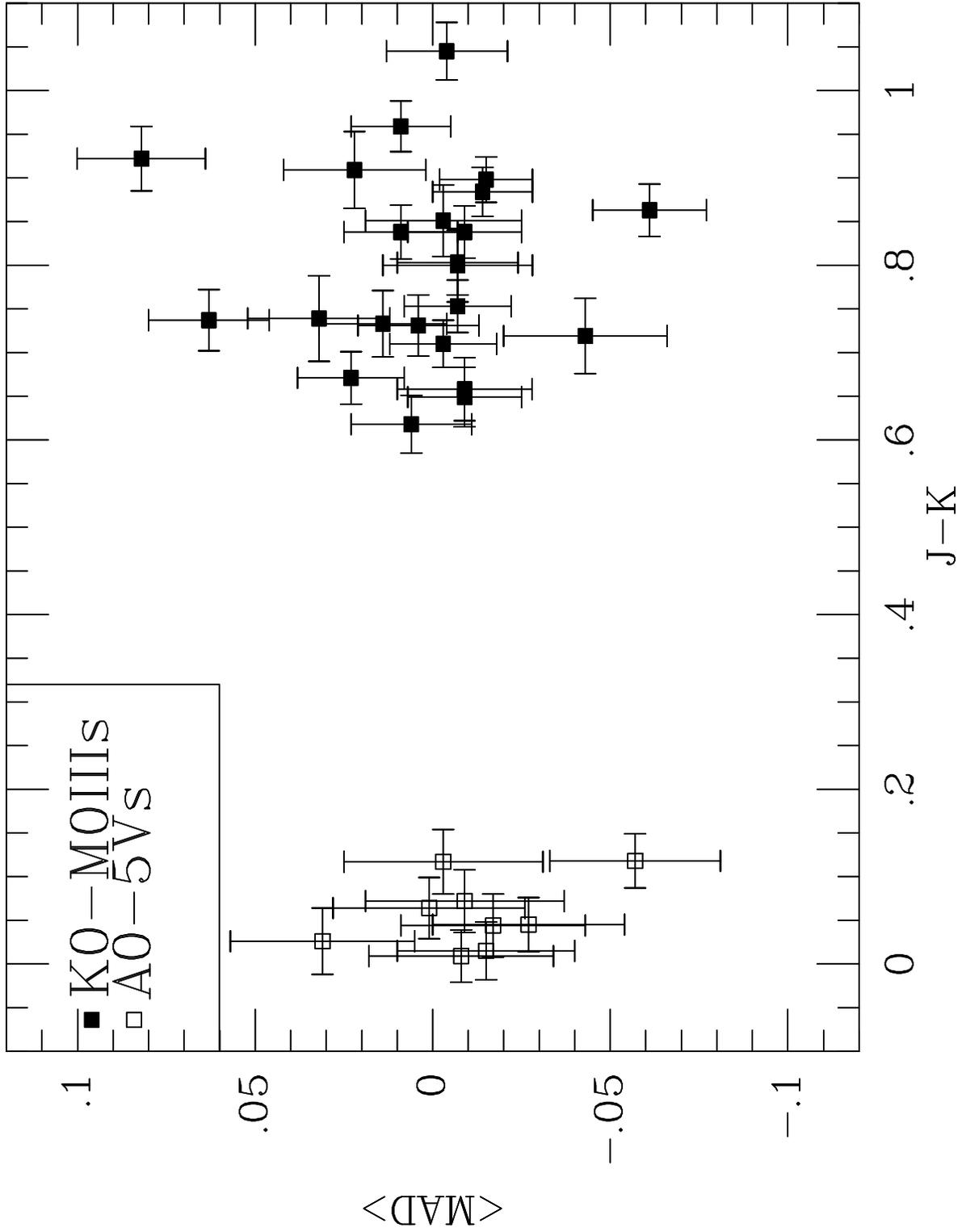}
\caption{As Figure~\ref{madspt} but plotted against 2MASS $J-K$ color.}
\label{madjk}
\end{figure}

\begin{figure}[tbp]
\centering
\vspace{20cm} \includegraphics{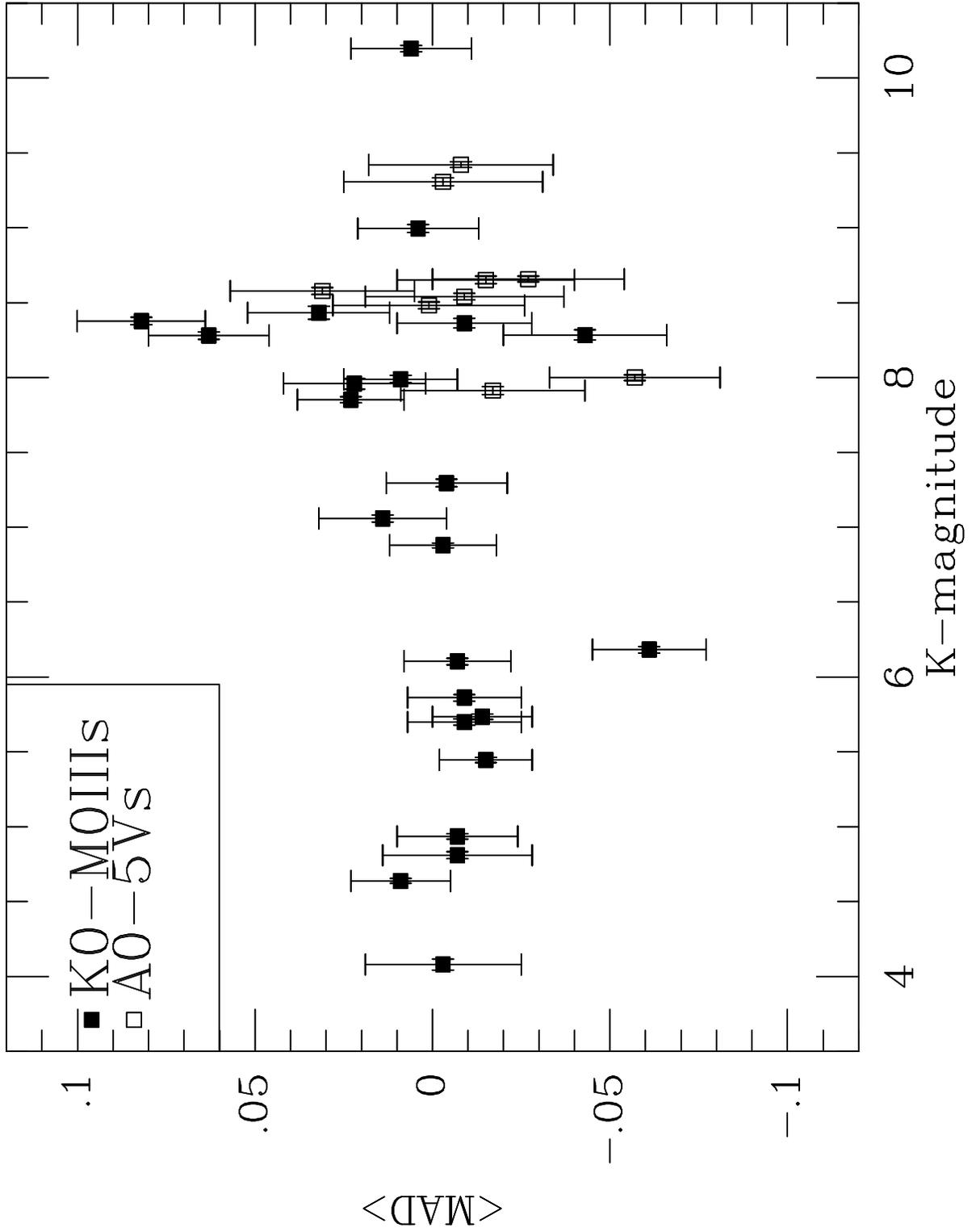}
\caption{As Figure~\ref{madspt} but plotted against 2MASS $K_s$ magnitude.} 
\label{madk}
\end{figure}

We have plotted the resulting MADs for the individual stars of the ensemble
against spectral type (Figure~\ref{madspt}), 
2MASS $J-K$ color (Figure~\ref{madjk}), and 2MASS $K$ magnitude (Figure~\ref{madk}) 
and found no perceptible biases with respect to these quantities.
The resulting ensemble-averaged ZPOs required to align 2MASS with our common
context by algebraically adding them to the final 2MASS magnitudes are: 
+0.001$\pm$0.005 ($J$); --0.019$\pm$0.007 ($H$); +0.017$\pm$0.005 ($K_{s}$).

\begin{table}[tbp]
\dummytable\label{cml}
\end{table}

\section{Conclusions}

We have defined element-by-element, photon-counting RSRs for all three 2MASS 
filter bands, incorporating the properties of detectors, filters, dichroics, 
lenses, coatings, dewar window, telescopes, and the earth's atmosphere. 
The RSRs are absolutely calibrated in
the common context of Cohen et al. (1999,2003), and zero point offsets have been
developed for 2MASS using thirty three new absolute calibrators of intermediate
brightness. 2MASS data are now directly supporting the development of faint
calibrators for IRAC and cross-calibrators between pairs of SIRTF
instruments using these RSRs to compute synthetic photometry.

\acknowledgments
We are grateful to Drs. Mike Skrutskie and Roc Cutri for providing us with the 
2MASS project's corporate memory and understanding of the nuances of the 
documentation of optical components available over the Web, and for valuable
comments on this manuscript.  We thank Dr. Leslie Hunt for supplying the
Rockwell generic HgCdTe DQE curve. 
This publication makes use of data products from the Two Micron All Sky 
Survey, which is a joint project of the University of Massachusetts and the 
Infrared Processing and Analysis Center/California Institute of Technology, 
funded by the National Aeronautics and Space Administration and the National 
Science Foundation.
MC's work on IRAC's calibrators is supported under contract SV9-69008 between
UC Berkeley and The Smithsonian Astrophysical Observatory. This research has
made use of the SIMBAD database, operated at CDS, Strasbourg, France.

\clearpage

\clearpage
\renewcommand{\arraystretch}{.6}

\clearpage

\begin{deluxetable}{lllllllll}
\rotate
\centering
\tablewidth{0pt} 
\tablenum{1}
\tablecolumns{9}
\tablecaption{Components of the 2MASS system that have been characterized. \label{instrument}}
\tablehead{Band& Telescope mirrors& Window& Camera lenses& $J$-dichroic& $H$-dichroic& Filter& DQE& Atmosphere\\}
\startdata
  $J$& primary/secondary& entrance& 7 with coatings& reflection& \nodata& $J$& generic& PLEXUS\\
  $H$& primary/secondary& entrance& 7 with coatings& transmission& reflection& $H$& generic& PLEXUS\\
$K_{s}$& primary/secondary& entrance& 7 with coatings& transmission& transmission& $K_{s}$& generic& PLEXUS\\
\enddata
\end{deluxetable}

\clearpage

\begin{deluxetable}{llllllll}
\rotate 
\centering
\tablewidth{0pt} 
\tablenum{2}
\tablecolumns{8}
\tablecaption{Zero magnitude attributes of 2MASS bands. \label{cal}}
\tablehead{Filter&  Bandwidth&  In-Band&  F$_{\lambda(iso)}$& $\lambda$(iso)&   Bandwidth&  F$_{\nu(iso)}$&  $\nu$(iso)\\   
& $\mu$m&  W cm$^{-2}$&   W cm$^{-2}$ $\mu$m$^{-1}$&  $\mu$m& Hz& Jy& Hz\\
& Uncertainty& Uncertainty& Uncertainty& Uncertainty& Uncertainty&  Uncertainty& Uncertainty\\
& $\mu$m&  \%&   W cm$^{-2}$ $\mu$m$^{-1}$&  $\mu$m& Hz& Jy& Hz\\}
\startdata
  $J$& 0.162&   5.082E-14& 3.129E-13& 1.235&  3.189E+13& 1594&  2.428E+14\\
&      0.001&   1.608&     5.464E-15& 0.006&  2.155E+11& 27.80& 2.746E+12\\
  $H$& 0.251&   2.843E-14& 1.133E-13& 1.662&  2.778E+13& 1024&  1.783E+14\\
 &     0.002&   1.721&     2.212E-15& 0.009&  2.540E+11& 19.95& 2.139E+12\\
$K_{s}$& 0.262& 1.122E-14& 4.283E-14& 2.159&  1.682E+13& 666.7& 1.390E+14\\
&      0.002&   1.685&     8.053E-16& 0.011&  1.409E+11& 12.55& 1.496E+12\\
\enddata
\end{deluxetable}

\clearpage

\begin{deluxetable}{lllll}
\centering
\tablewidth{0pt} 
\tablenum{3}
\tablecolumns{5}
\tablecaption{Zero magnitude attributes of 2MASS $J$-band under
different water vapor conditions. \label{water}}
\tablehead{Water&  Bandwidth&  In-Band&  F$_{\lambda(iso)}$& $\lambda$(iso)\\   
mm&     $\mu$m&    W cm$^{-2}$&   W cm$^{-2}$ $\mu$m$^{-1}$& $\mu$m\\
& Uncertainty& Uncertainty& Uncertainty& Uncertainty\\
& $\mu$m&  W cm$^{-2}$&   W cm$^{-2}$ $\mu$m$^{-1}$&  $\mu$m\\}
\startdata
5&  0.170& 5.319E-14& 3.127E-13& 1.235\\
& 0.001& 8.507E-16& 2.901E-15& 0.003\\
0.5& 0.192& 5.888E-14& 3.060E-13& 1.243\\
& 0.001& 9.321E-16& 2.638E-15& 0.003\\
\enddata
\end{deluxetable}
 
\clearpage

\begin{deluxetable}{lllrrrrrrrrrr}
\renewcommand{\arraystretch}{.6} 
\tabletypesize{\tiny}
\rotate
\centering
\tablewidth{0pt} 
\tablenum{4}
\tablecolumns{13}
\tablecaption{Predicted and observed 2MASS $JHK_{s}$ for Landolt and Carter-Meadows stars. Magnitudes
and differences marked by asterisks were rejected as discussed in the text. \label{cml}}
\tablehead{HD& SA/SAO& Type& Predicted~$J$& Observed~$J$& Difference& Predicted~$H$& Observed~$H$& Difference& Predicted~$K_{s}$& Observed~$K_{s}$& Difference& MAD\\}
\startdata 
  HD5319& SA92-336&  K0III&   6.323$\pm$0.011& 6.348$\pm$0.027& +0.025$\pm$0.029& 5.838$\pm$0.009& 5.864$\pm$0.044&  +0.026$\pm$0.045& 5.735$\pm$0.008& 5.699$\pm$0.020& --0.036$\pm$0.022& +0.005\\
 \nodata& SA94-251&  K1III&   9.063$\pm$0.020& 9.003$\pm$0.027& --0.060$\pm$0.034& 8.459$\pm$0.018& 8.425$\pm$0.051& --0.034$\pm$0.054& 8.311$\pm$0.016& 8.284$\pm$0.033& --0.027$\pm$0.037& --0.040\\
 \nodata& SA103-526& K0III&   9.016$\pm$0.016& 9.020$\pm$0.019&  +0.004$\pm$0.025& 8.506$\pm$0.013& 8.475$\pm$0.053& --0.031$\pm$0.055& 8.386$\pm$0.013& 8.362$\pm$0.031& --0.024$\pm$0.033& --0.017\\
HD118280& SA105-205& K3III&   6.366$\pm$0.010& 6.343$\pm$0.020& --0.023$\pm$0.023& 5.703$\pm$0.008& 5.726$\pm$0.034&  +0.023$\pm$0.035& 5.466$\pm$0.007& 5.445$\pm$0.017& --0.021$\pm$0.018& --0.007\\
HD118290& SA105-405& K5III&   5.615$\pm$0.009& 5.597$\pm$0.024& --0.018$\pm$0.026& 4.851$\pm$0.007& 4.879$\pm$0.059&  +0.028$\pm$0.059& 4.619$\pm$0.006& 4.638$\pm$0.016&  +0.019$\pm$0.017& +0.010\\
HD139308& SA107-35&  K2III&   5.607$\pm$0.011& 5.609$\pm$0.034&  +0.002$\pm$0.036& 4.935$\pm$0.009& $^*$5.058$\pm$0.040&  $^*$0.123$\pm$0.041& 4.821$\pm$0.009& 4.809$\pm$0.024& --0.012$\pm$0.026& --0.005\\
HD139513& SA107-347& K1.5III& 7.058$\pm$0.017& 7.045$\pm$0.021&  --0.013$\pm$0.027& 6.414$\pm$0.010& 6.328$\pm$0.034& --0.086$\pm$0.035& 6.270$\pm$0.009& 6.182$\pm$0.022& --0.088$\pm$0.024& --0.062\\
 \nodata& SA107-484& K3III&   9.151$\pm$0.014& 9.170$\pm$0.021&  +0.019$\pm$0.025& 8.525$\pm$0.012& 8.512$\pm$0.042&  --0.013$\pm$0.044& 8.311$\pm$0.010& 8.431$\pm$0.044&  +0.120$\pm$0.045& +0.042\\
 \nodata& SA108-475& K3III&   8.845$\pm$0.015& 8.828$\pm$0.019& --0.017$\pm$0.024& 8.174$\pm$0.012& 8.148$\pm$0.036& --0.026$\pm$0.038& 7.933$\pm$0.010& 7.990$\pm$0.024&  +0.057$\pm$0.026& +0.005\\
HD149845& SA108-827& K2III&   5.744$\pm$0.012& 5.738$\pm$0.032& --0.006$\pm$0.034& 5.062$\pm$0.010& $^*$5.187$\pm$0.017&  $^*$0.125$\pm$0.020& 4.943$\pm$0.009& 4.935$\pm$0.018& --0.008$\pm$0.020& --0.007\\
 \nodata& SA108-1918&K3III&   8.840$\pm$0.015& 8.868$\pm$0.025& +0.028$\pm$0.029& 8.154$\pm$0.012& 8.128$\pm$0.038& --0.026$\pm$0.040& 7.903$\pm$0.011& 7.959$\pm$0.036&  +0.056$\pm$0.038& +0.019\\
 \nodata& SA109-231& K2III&   6.746$\pm$0.015& 6.700$\pm$0.021& --0.046$\pm$0.026& 6.012$\pm$0.012& 6.050$\pm$0.033&  +0.038$\pm$0.035& 5.862$\pm$0.011& 5.862$\pm$0.022&  +0.000$\pm$0.025& --0.003\\
HD172651& SA110-471& K2III&   4.933$\pm$0.011& 4.930$\pm$0.019&  --0.003$\pm$0.022& 4.204$\pm$0.009& $^*$3.876$\pm$0.220& $^*$-0.328$\pm$0.220& 4.055$\pm$0.008& $^*$4.079$\pm$0.036&  $^*$0.024$\pm$0.037& --0.003\\
 \nodata& SA112-275& K0III&   7.749$\pm$0.015& 7.791$\pm$0.029&  +0.042$\pm$0.033& 7.199$\pm$0.012& 7.197$\pm$0.036& --0.002$\pm$0.038& 7.055$\pm$0.011& 7.058$\pm$0.024& +0.003$\pm$0.026& +0.014\\
 \nodata& SA112-595& M0III&   8.361$\pm$0.014& 8.341$\pm$0.021& --0.020$\pm$0.025& 7.486$\pm$0.011& 7.502$\pm$0.042&  +0.016$\pm$0.043& 7.289$\pm$0.010& 7.296$\pm$0.026& +0.007$\pm$0.028& +0.001\\
 \nodata& SA113-259& K2III&   9.745$\pm$0.018& 9.725$\pm$0.023& --0.020$\pm$0.029& 9.099$\pm$0.016& 9.132$\pm$0.021&  +0.033$\pm$0.027& 9.002$\pm$0.016& 8.994$\pm$0.026& --0.008$\pm$0.030& +0.001\\
 \nodata& SA113-269& K0III&   7.548$\pm$0.014& 7.589$\pm$0.021&  +0.041$\pm$0.025& 7.030$\pm$0.011& 7.010$\pm$0.042&  --0.020$\pm$0.043& 6.906$\pm$0.010& 6.879$\pm$0.017& --0.027$\pm$0.020& --0.002\\
HD215141& SA114-176& K4III&   6.673$\pm$0.014& 6.618$\pm$0.021& --0.055$\pm$0.025& 5.942$\pm$0.012& 5.946$\pm$0.029&  +0.004$\pm$0.031& 5.729$\pm$0.010& 5.734$\pm$0.018& +0.005$\pm$0.021& --0.015\\
 \nodata& SA114-548& K3III&   9.187$\pm$0.015& 9.300$\pm$0.026&  +0.113$\pm$0.030& 8.524$\pm$0.012& 8.528$\pm$0.040&  +0.004$\pm$0.042& 8.287$\pm$0.011& 8.378$\pm$0.026& +0.091$\pm$0.028& +0.069\\
 \nodata& SA114-656& K1III&  10.841$\pm$0.022&10.815$\pm$0.026& --0.026$\pm$0.034& 10.290$\pm$0.020& 10.301$\pm$0.021&  +0.011$\pm$0.029& 10.174$\pm$0.019& 10.197$\pm$0.021& +0.023$\pm$0.028& +0.003\\
 \nodata& SA114-670& K1.5III&   8.950$\pm$0.020& 9.016$\pm$0.024& +0.066$\pm$0.029& 8.336$\pm$0.013& 8.390$\pm$0.027& +0.054$\pm$0.031& 8.210$\pm$0.012& 8.279$\pm$0.026& +0.069$\pm$0.028& +0.063\\
HD222732& SA115-427& K2III&   6.860$\pm$0.013& 6.857$\pm$0.021& --0.003$\pm$0.025& 6.216$\pm$0.011& 6.214$\pm$0.026& --0.002$\pm$0.028& 6.119$\pm$0.011& 6.104$\pm$0.022& --0.015$\pm$0.025& --0.007\\
 \nodata& SA115-516& K1.5III& 8.506$\pm$0.018& 8.522$\pm$0.021&  +0.016$\pm$0.027& 7.927$\pm$0.011& 7.948$\pm$0.024& +0.021$\pm$0.027& 7.822$\pm$0.011& 7.851$\pm$0.021& +0.029$\pm$0.024& +0.022\\
HD197806&   \nodata& K0III&   7.563$\pm$0.014& 7.585$\pm$0.024& +0.022$\pm$0.028& 7.022$\pm$0.012& 7.027$\pm$0.044& +0.005$\pm$0.046& 6.884$\pm$0.011& 6.849$\pm$0.031& --0.035$\pm$0.033& --0.002\\
 HD15911& SAO232803& A0V&     9.439$\pm$0.035& 9.430$\pm$0.023& --0.009$\pm$0.042& 9.483$\pm$0.041& 9.497$\pm$0.023& +0.014$\pm$0.047& 9.450$\pm$0.043& 9.421$\pm$0.019& --0.029$\pm$0.047& --0.008\\
 HD29250& SAO169590& A4V&     9.419$\pm$0.036& 9.425$\pm$0.026& +0.006$\pm$0.045& 9.386$\pm$0.041& 9.383$\pm$0.027& --0.003$\pm$0.049& 9.322$\pm$0.042& 9.308$\pm$0.026& --0.014$\pm$0.050& --0.003\\
 HD62388& SAO153304& A0V&    8.715$\pm$0.033& 8.702$\pm$0.025& --0.013$\pm$0.042& 8.734$\pm$0.038& 8.684$\pm$0.040& --0.050$\pm$0.055& 8.686$\pm$0.040& 8.657$\pm$0.019& --0.029$\pm$0.044& --0.030\\
 HD71264& SAO135911& A0V&    8.572$\pm$0.034& 8.603$\pm$0.030& +0.031$\pm$0.045& 8.572$\pm$0.038& 8.571$\pm$0.024& --0.001$\pm$0.045& 8.512$\pm$0.040& 8.577$\pm$0.023& +0.065$\pm$0.046& +0.032\\
 HD84090& SAO221405& A3V&    8.534$\pm$0.032& 8.546$\pm$0.027& +0.012$\pm$0.042& 8.531$\pm$0.037& 8.500$\pm$0.047& --0.031$\pm$0.060& 8.477$\pm$0.038& 8.482$\pm$0.023& +0.005$\pm$0.045& --0.004\\
HD105116& SAO223215& A2V&    8.143$\pm$0.032& 8.117$\pm$0.024& --0.026$\pm$0.040& 8.126$\pm$0.036& 8.033$\pm$0.027& --0.093$\pm$0.045& 8.060$\pm$0.037& 7.999$\pm$0.020& --0.061$\pm$0.042& --0.060\\
HD106807& SAO223331& A1V&    8.698$\pm$0.033& 8.667$\pm$0.021& --0.031$\pm$0.040& 8.704$\pm$0.038& 8.693$\pm$0.026& --0.011$\pm$0.046& 8.648$\pm$0.039& 8.652$\pm$0.025& +0.004$\pm$0.047& --0.013\\
HD136879& SAO253162& A0V&    8.615$\pm$0.035& 8.613$\pm$0.029& --0.002$\pm$0.045& 8.620$\pm$0.039& 8.556$\pm$0.045& --0.046$\pm$0.060& 8.535$\pm$0.040& 8.541$\pm$0.021& +0.006$\pm$0.045& --0.014\\
HD216009& SAO231319& A0V&    7.961$\pm$0.030& 7.957$\pm$0.024& --0.004$\pm$0.038& 7.988$\pm$0.034& 7.966$\pm$0.042& --0.022$\pm$0.054& 7.945$\pm$0.036& 7.913$\pm$0.027& --0.032$\pm$0.045& --0.020\\
\enddata
\end{deluxetable}

\end{document}